# A Novel Microdata Privacy Disclosure Risk Measure

**Marmar Orooji and Gerald M. Knapp**
**Louisiana State University**
**Baton Rouge, LA 70803, USA**

## Abstract

A tremendous amount of individual-level data is generated each day, of use to marketing, decision makers, and machine learning applications. This data often contain private and sensitive information about individuals, which can be disclosed by "adversaries". An adversary can recognize the underlying individual's identity for a data record by looking at the values of quasi-identifier (QID) attributes, known as identity disclosure, or can uncover sensitive information about an individual through attribute disclosure. In Statistical Disclosure Control, multiple disclosure risk measures have been proposed. These share two drawbacks: they do not consider identity and attribute disclosure concurrently in the risk measure, and they make restrictive assumptions on an adversary's knowledge by assuming certain attributes are QIDs and there is a clear boundary between QIDs and sensitive information. In this paper, we present a novel disclosure risk measure that addresses these limitations, by presenting a single combined metric of identity and attribute disclosure risk, and providing flexibility in modeling adversary's knowledge. We have developed an efficient algorithm for computing the proposed risk measure and evaluated the feasibility and performance of our approach on a real-world dataset from the domain of social work.

## Keywords
Privacy Preserving, Privacy Disclosure Attack, Risk Assessment

## 1. Introduction and Background

Individual-level data (aka microdata) is of interest to researchers and decision makers in many domains, for tracking, classifying, and predicting people's behavior. However, it often contains private and sensitive information about individuals, such as Electronic Health Records, and thereby cannot be made freely available for public access. Thus, the question is how data owners can share their data for research purposes while not violating individuals' privacy. *Statistical Disclosure Control (SDC)* and *Privacy Preserving Data Publishing (PPDP)* are two areas in privacy field that have recently studied this problem in depth.

For privacy preservation, direct identifiers that are unique to a person, such as social security number and phone number, are removed from the microdata. However, de-identification is not guaranteed because of the existence of quasi-identifiers (QIDs). QIDs are attributes, which are not individually unique to a person, but when considered as a set the combination of QID field values can be used to identify individuals with high probability. Examples are birthdate, gender, and zip code. Another class of attributes in microdata are Sensitive Attributes (SA), which contain sensitive information about individuals such as those relating to health, education, criminal history, or income.

An "adversary" is a person who wants to disclose sensitive information about a person, called the "victim". The adversary typically knowns the values of the QIDs of the victim from personal knowledge or publicly available datasets (local census data, voter lists, social media, …), and uses this information to match against QIDs appearing in the confidential dataset. For example Sweeney [1] demonstrated discovering the medical record of the governor of Massachusetts from data released by the Group Insurance Commission, after obtaining the QID-values for the Governor from a public voter registration list. Sweeney notes that 87% of U.S. citizens can be uniquely recognized in datasets using only their birth date, gender, and 5-digit zip code. Although QIDs might lead to disclosure, they generally cannot be removed from the dataset since they carry useful information for research and analysis purposes.

Two main disclosure types that have been studied in depth in the literature are *Identity Disclosure* and *Attribute Disclosure. Identity Disclosure* occurs when an adversary can recognize that a record in the released dataset belongs to an individual by matching QID values. *Attribute Disclosure* occurs when an adversary can link a sensitive value to an individual. In the latter, the adversary may not precisely identify a record of a specific victim, but could infer his/her sensitive values from the published data, based on the set of sensitive values associated with the group that the victim belongs to.

Previous studies in SDC have proposed techniques to measure the risk of identity disclosure. These risk measures are defined based on either uniqueness or re-identification. In uniqueness measures, risk is defined as the probability that the rare combination of QID values in the privacy preserved dataset is indeed rare in the population dataset [2]. Re-identification methods estimate the number of re-identifications an adversary can obtain by matching QIDs from his external knowledge against confidential dataset through record linkage algorithms [3, 4]. These methods require the assumption of knowing the exact external knowledge for adversary. Domingo-Ferrer addressed this issue by proposing the "maximum knowledge attacker model", which considers an adversary who knows the values of all QIDs in the confidential dataset about a victim [5]. This model presents the maximum disclosure risk.

Although identity disclosure risk measures have been studied in depth, the literature lacks measuring attribute disclosure attacks. In PPDP, various privacy models for attribute disclosure have been proposed. However, they are not providing a measurable condition – i.e., they are not measuring the risk of attribute disclosure of a dataset; instead, they are specifying a Boolean condition in which the dataset is prevented from attribute disclosure if it satisfies the condition [6]. For instance, Machanavajjhala et al. proposed a privacy model named "$\ell$-Diversity", which requires the records with similar values in their QIDs have diverse sensitive values [7]. As another privacy model, "t-closeness" requires the distribution of sensitive values in each group of records with similar QID values to be close to the overall distribution [8].

There are two common drawbacks among privacy techniques in SDC and PPDP. First, they restrict an adversary's knowledge by considering a specific number of attributes as QIDs. However, in real-world applications, many adversaries exist with different external knowledge. Addressing this issue, Motwani and Xu proposed distinct ratio and separation ratio measures to find QIDs [9]. These measures are defined based on value frequencies in different combinations of attribute such that the combinations that lead to more unique values are more likely to form QIDs. Secondly, they consider a clear boundary between the set of QIDs and SAs. Nevertheless, in practice, an attribute can be a QID and SA simultaneously – e.g., *Occupation*.

In this paper, we propose a novel record-level disclosure risk measure, which considers both identity and attributes disclosure attack concurrently. In addition, it provides flexibility in modeling adversary's knowledge by assigning likelihoods of data attributes being publicly known instead of predefining specific attributes as QIDs, considering any combination of attributes being known by an adversary, and allowing QID and SA attributes to have overlap with each other. We also provide a mechanism for assigning different sensitivity weights for SAs as well as specific values of SAs. This can improve modeling and control of attribute disclosure.

## 2. Methodology

### 2.1 Novel Disclosure Risk Measure

Our disclosure risk measure is defined at record level, based on risk assessment methodology, where risk of disclosure equals likelihood times consequence. Likelihood of a record $r$ indicates the likelihood of it being identified by an adversary, and captures identity disclosure attack in risk measurement. Consequence specifies that given *r* is identified, what is the sensitivity of the private information of *r* being revealed? This term captures attribute disclosure attack.

In calculating risk, we iteratively split attributes into a known set and unknown set of attributes. A known set contains attributes, which an adversary knows about a victim (QIDs). The remaining attributes form the unknown set and are attributes, which may contain private information (SAs) an adversary wants to disclose about a victim. The unknown set is a complement set of the known set. Likelihood is calculated based on the known set and consequence based on the unknown set. All subsets of possible known sets and their compliment unknown subsets are considered. Thus, we are allowing an attribute to contribute in both known and unknown set and thereby allowing the overlap between QIDs and SAs. Since each attribute has 2 possibilities − i.e., being in the known set or in unknown set - given $m$ attributes, the total number of known-unknown sets is equal to $2^m$. Our proposed disclosure risk measure is then defined as:

$$D(r) = \sum_{i=1}^{2^m} L_{KS_i}(r) \times \alpha\, C_{UKS_i}(r) \tag{1}$$

where D is the disclosure risk, L is likelihood, C consequence, $KS_i$ is the $i^{th}$ known set, $UKS_i$ is the $i^{th}$ unknown set, and $\alpha$ is a consequence coefficient which is a constant value larger than 1.

$L_{KS_i}(t)$ is the likelihood of identifying the record $r$ through the attributes in the $i^{th}$ known set. It comprises two terms – i.e., the probability of the $i^{th}$ known set being publicly known, and the uniqueness of the $i^{th}$ known set's attribute values for the record $r$ in the whole dataset. It is formulated as:

$$L_{KS_i}(r) = PK(KS_i) \times \frac{1}{count\left(r(KS_i)\right)} \tag{2}$$

$$PK(KS_i) = \prod_{A_j \in KS_i} PK\left(A_j\right) \tag{3}$$

where $PK(KS_i)$ is the probability of $KS_i$ being publicly known, $A_j$ is the $j^{th}$ attribute in $KS_i$, $r(KS_i)$ is $KS_i$'s attribute values for the record $r$, and $count\left(r(KS_i)\right)$ is the number of occurrences of $r(KS_i)$ in the dataset.

The first term in the likelihood expression is the probability of the $i^{th}$ known set being publicly known. To make calculations tractable, it is assumed that the likelihood of publicly knowing each attribute is independent from one another. The probability of $KS_i$ is then simply the multiplication of the publicly known probabilities of each attribute in $KS_i$. These probabilities are assigned by data publishers; they have the flexibility to choose probabilities such that best considers adversaries' knowledge.

A known set can be very likely to be publicly known but the sets attribute values for a record might occur frequently in the dataset. Thus, an adversary will find several matched records for the victim and therefore the identity disclosure likelihood is decreased. Hence, the second term considers the values of the known set of attributes for the record $r$. Then it counts the number of occurrences of those values together in the whole dataset. If this count is large, it means a large numbers of records have these values and the person whom record $r$ belongs to is less likely to be identified by those attributes. Therefore, the count is inversely correlated with likelihood. Including this term, likelihood is no longer a probability function. However, for each record it is still a value between 0 and 1.

$C_{UKS_i}(t)$ is the consequence of identifying record $r$ based on the $i^{th}$ unknown set, which means how much private information of record $r$ is revealed, after being identified. In order to measure the level of private information of record $r$, which is stored in the unknown set, data publisher needs to assign sensitivity weights to both attributes and their values. Accordingly, consequence of record $r$ based on the $i^{th}$ unknown set is derived as:

$$C_{UKS_i}(r) = \sum_{A_j \in UKS_i} W\left(A_j\right) \times W\left(r\left(A_j\right)\right) \tag{4}$$

where $r\left(A_j\right)$ is the value of $A_j$ for record $r$, $W\left(A_j\right)$ is the sensitivity weight of $A_j$, and $W\left(r\left(A_j\right)\right)$ is the sensitivity weight of $r\left(A_j\right)$. The sensitivity weight implies how much the information is sensitive and private to be disclosed about a person. Data publishers have the flexibility to assign weights based on perceived sensitivity of the data. The assigned weights are between 0 and 1.

Table 1. Sample microdata

| | Age | Gender | Race | Income | Disease |
|---|---|---|---|---|---|
| $r_1$ | 34 | Male | Black | 60K | Flu |
| $r_2$ | 19 | Female | White | 36K | Flu |
| $r_3$ | 40 | Male | Asian-Pac-Islander | 45K | Flu |
| $r_4$ | 34 | Male | Black | 50K | Cancer |
| $r_5$ | 51 | Female | Black | 65K | Flu |

As an example, consider a microdata shown in **Table 1**. We assigned probabilities and sensitivity weights as shown in **Table 2**. Considering one split of attributes {age, gender, race} as the known set and {income, disease} as the unknown set, likelihood and consequence for $r_4$ is calculated as:

$$L_{\{age,gender,race\}}(r_4) = (0.3 \times 0.8 \times 0.7\ ) \times (1/2) = \ 0.084$$

$$C_{\{income,disease\}}(r_4) = (0.9 \times 0.7) + (1 \times 1) = 1.63$$

Table 2. Assigned probabilities and sensitivity weights for sample microdata

| Attribute | Publicly Known Probability | Attribute Sensitivity Weight | Value Sensitivity Weight | |
|---|---|---|---|---|
| | | | Value | Weight |
| **Age** | 0.3 | 0 | ---------------------- | |
| **Gender** | 0.8 | 0 | ---------------------- | |
| **Race** | 0.7 | 0 | ---------------------- | |
| **Income** | 0.005 | 0.9 | Less than 40K | 1 |
| | | | Between 40K and 70K | 0.7 |
| **Disease** | 0.001 | 1 | Flu | 0.2 |
| | | | Cancer | 1 |

### 2.2 Calculation Efficiency

Our proposed disclosure risk measure is calculated over all possible known/unknown sets of attributes – i.e., $2^m$ number of sets where $m$ is the number of attributes in the dataset. Calculating our risk measure can be computationally expensive in datasets with large number of attributes, due to the exponential growth in the number of known/unknown sets. To make our risk measure computationally feasible, we reduce the number of known/unknown sets by pruning known sets, which lead to low value of likelihood and thereby low contribution in risk measure. If a known subset of attributes is highly unlikely to be known, any larger known set containing this subset is even more unlikely, due to the product term of probabilities. Thus, if publicly known probability of a known set is less than a threshold $\varepsilon$, any superset of that known set will also have the probability of less than $\varepsilon$. Consider a tree of all subsets of known attributes; each node represents a subset. The root is the empty subset and at each level of the tree, the size of the subsets grows by one. Since $PK(KS_i)$ of node *i* monotonically decreases as we traverse the tree from the root to the leaves, by pre-order tree traversal, we can prune the subtree at each node with $PK(KS_i)$ less than $\varepsilon$.

<u>Pruning Algorithm</u>

```
SUB ={{}}; NEXT ={{1,2,3,…,m }}; i=0  // initialization
while ( (length(SUB)-i) ≠ 0 ) {          // loop until no subsets are available for checking their probability
      sub = SUB ( length(SUB)-i );       // select a subset at which the pointer is pointing
      nxt = NEXT ( length(NEXT)-i );     // select the corresponding remaining attributes
      while ( length(nxt) ≠ 0 ) {        // loop until no attributes remained for expanding the current subset
            U = sub ∪ first attribute index in nxt;
            Remove first index in nxt and update nxt;
            NEXT ( length(NEXT)-i ) = nxt;
            if ( product of probabilities of attributes in U ) > ε){
                  Append U to SUB;
                  Append a set containing indices larger than the maximum index in U  to NEXT;
                  Select last set in SUB as sub and last set in NEXT as nxt;
                  Reset i to 0;
            }
      }
      Increment i by one;
}
```

Figure 1. Algorithm for creating known sets with publicly known probability of more than $\varepsilon$

Figure 1 summarizes the pruning algorithm, which calculates our proposed disclosure risk measure feasible. In this algorithm, we refer to attributes by their indices (1,2,…, m). *SUB* is a list, which stores the subsets of attributes with publicly known probability more than $\varepsilon$ and initially it contains empty subset. Each subset in *SUB* has a corresponding element in *NEXT,* which shows the remaining attributes that can be appended to that subset. Thus, *NEXT* is initiated by {1,2,…, m} which are attribute indices that can be appended to an empty subset. *i* is a pointer sweeping *SUB* and *NEXT* to point to the current subset and select it for checking whether it needs to be enlarged or pruned. After a subset is selected, its remaining attributes shown in a corresponding set in *NEXT* should be appended one by one to the selected subset and each time the probability of the new subset must be checked. If it is more than $\varepsilon$, the new enlarged subset is appended to *SUB* and indices larger than the maximum index in the enlarged subset are appended to *NEXT* as a set. Then $i$ is reset to 0 to start sweeping *SUB* and *NEXT* from the end. If not, the new subset is not appended to *SUB* and will not be enlarged. When no more attributes remained for the selected subset, $i$ is incremented by one to select the prior subset in *SUB* and its corresponding remaining attributes in *NEXT*.

## 3. Experiment

We have implemented our approach on a sensitive dataset provided by the Social Research and Evaluation Center at Louisiana State University. The dataset contains data from Louisiana Department of Education (LADOE), Office of Juvenile Justice (LAOJJ), and Department of Corrections (LADOC). This dataset contains 1009993 students enrolled in Louisiana public schools between 1999-2011 school years. There are 1,009,993 records (one record per student), and 27 attributes. Examples of attributes are age, gender, ethnicity, homelessness, dropout flag, LADOC contact, and LAOJJ contact. This dataset carries valuable information for research purposes in social work domain [10].

To calculate disclosure risk of each record based on our proposed measure, we need to assign publicly known probabilities and sensitivity weights to each attribute, sensitivity weights to the values within each attribute, and consequence coefficient (α). The sensitivity weights assigned are either 0 or 1. We set α to be 100 to magnify the effect of consequence. Examples of the assigned probabilities and weights for sample attributes are shown in Table 3.

Table 3. Examples of assigned publicly known probabilities and sensitivity weights

| **Attribute** | **Publicly Known Probability** | **Attribute Sensitivity Weight** | **Value Sensitivity Weight** | |
|---|---|---|---|---|
| | | | **Value** | **Weight** |
| **Age** | 0.05 | 0 | --------- | |
| **Gender** | 0.8 | 0 | --------- | |
| **Dropout Flag** | 0.005 | 1 | Yes | 1 |
| | | | No | 0 |
| **LADOC Contact Flag** | 0.001 | 1 | Yes | 1 |
| | | | No | 0 |

Since there are 27 attributes, initially there will be $2^{27} = 134{,}217{,}728$ known/unknown sets. We employed our pruning algorithm to reduce the number of subsets. In this experiment, we assigned the threshold $\varepsilon$ to be 0.01, which implies known sets with publicly known probabilities less than 1% are pruned. After applying our proposed pruning algorithm, only 111 subsets of attributes remained. Thus, we calculated the disclosure risk of each record based on our proposed measure over 111 known and unknown sets. Having all parameters being specified, we have calculated the disclosure risk measure for each record in the dataset and the histogram of risk values among 1,009,993 records is shown in Figure 2. It is illustrated that 15,651 records (1.55%) have risk value more than 0.01. To further decrease this number of high risk records, we can apply anonymization only on the high risk records and decrease their disclosure risk [11].

## 4. Conclusions

Studying privacy disclosure risk measures in the literature, no single metric has been proposed to measure the risk of both identity and attribute disclosure attack at the same time. We have formulated both types of disclosure in a single combined metric, which can be measured for each record in a dataset. Our proposed disclosure risk measure is calculated based on the parameters, which are assigned by a data publisher according to the nature of the dataset. Data publisher needs to determine the probability of being publicly known for each attribute and the sensitivity level of each attribute and each value within an attribute in terms of containing private information. These parameters improve modeling disclosure attacks, compared to the previous works in literature where data publishers need to strictly determine specific number of QIDs and SAs, with no overlap. In addition, we are modeling adversaries with any possible background knowledge by considering any combination of attributes to be known by adversary, contrary to previous works where they restrict adversaries' knowledge. In order to make the calculation of our risk measure feasible for datasets with large number of attributes, we have developed a pruning algorithm and evaluated the feasibility of our approach through an experiment on real-world sensitive dataset.

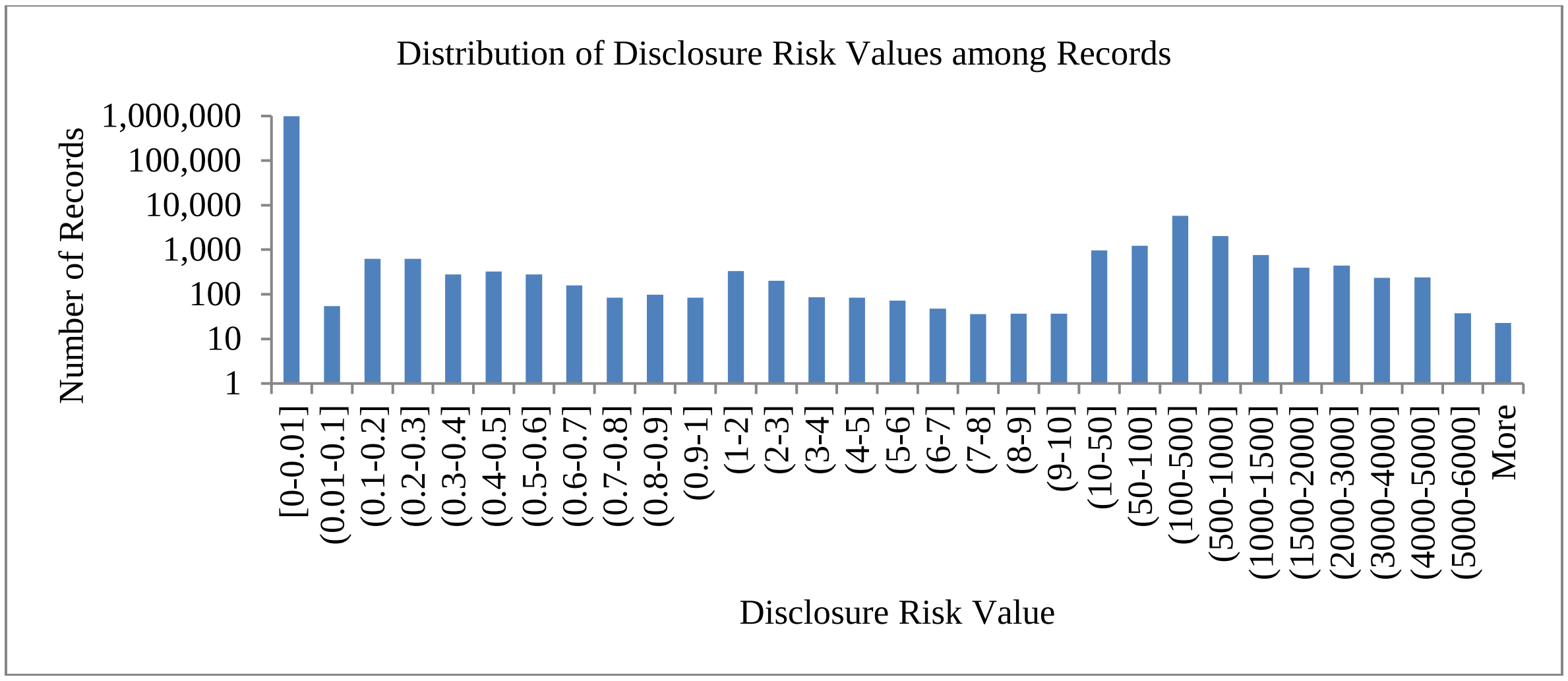

Figure 2. Histogram of Records Disclosure Risk Values (Logarithmic Scale)

Besides the usability of our measure in privacy disclosure risk assessment, it can be useful for improving privacy preserving data publishing techniques. These techniques can be optimized and better preserve data utility if they get modified to target records with high disclosure risk.

## Acknowledgements

The authors would like to acknowledge the Social Research and Evaluation Center at Louisiana State University for providing this research a real-world sensitive dataset in social work domain.